\documentclass{article}

\usepackage{arxiv}

\usepackage[utf8]{inputenc} % allow utf-8 input
\usepackage[T1]{fontenc}    % use 8-bit T1 fonts
\usepackage{hyperref}       % hyperlinks
\usepackage{url}            % simple URL typesetting
\usepackage{booktabs}       % professional-quality tables
\usepackage{amsfonts}       % blackboard math symbols
\usepackage{amsmath}        % equations
\usepackage{nicefrac}       % compact symbols for 1/2, etc.
\usepackage{microtype}      % microtypography
\usepackage{lipsum}	     	% Can be removed after putting your text content
\usepackage{graphicx}
\usepackage{natbib}
\usepackage{doi}
\usepackage[subpreambles=true]{standalone}
\usepackage{import}
\usepackage[ruled,vlined,english]{algorithm2e}
\usepackage{tabularx}

\newcommand{\expnumber}[2]{\texttt{{#1}e{#2}}} % command for 1e-10 etc.

\title{Simulation-Driven COVID-19 Epidemiological Modeling with Social Media}

\date{June 22, 2021}	% Here you can change the date presented in the paper title
\author{ \href{https://orcid.org/0000-0002-0559-5176}{\includegraphics[scale=0.06]{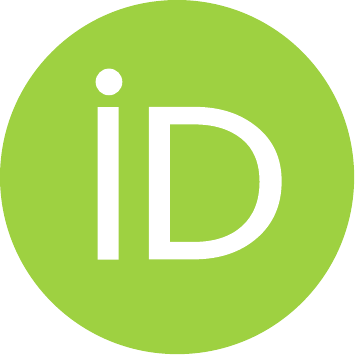}\hspace{1mm}Jose Storopoli}\thanks{We would to thank João Vinicíus, Elias, Paula, Camila and Leandro for annotating the tweets.} \\
	Department of Computer Science\\
	Universidade Nove de Julho - UNINOVE\\
	São Paulo, Brazil \\
	\texttt{josees@uni9.pro.br} \\
	\And 
	\href{https://orcid.org/0000-0002-5232-0781}{\includegraphics[scale=0.06]{orcid.pdf}\hspace{1mm}André Luís Marques Ferreira dos Santos} \\
	Department of Business Administration\\
	Universidade Nove de Julho - UNINOVE\\
	São Paulo, Brazil \\
	\texttt{andrelmfsantos@gmail.com} \\
	\And
	\href{https://orcid.org/0000-0002-3799-9415}{\includegraphics[scale=0.06]{orcid.pdf}\hspace{1mm}Alessandra Cristina Guedes Pellini} \\
	Medical School \\
    Universidade Nove de Julho - UNINOVE\\
	São Paulo, Brazil \\
	\texttt{alessandra.pellini@uni9.pro.br} \\
	\And
	Breck Baldwin \\
	\texttt{breckbaldwin@gmail.com} \\
}

\renewcommand{\shorttitle}{Simulation-Driven COVID-19 Epidemiological Modeling with Social Media}

\hypersetup{
pdftitle={Simulation-Driven COVID-19 Epidemiological Modeling with Social Media},
pdfsubject={[q-bio, stat]},
pdfauthor={Jose~Storopoli et al},
pdfkeywords={COVID, Bayesian Modeling, Epidemiology, Simulation, Social Media, Twitter},
}

\begin{document}
\maketitle

\begin{abstract}
	Modern Bayesian approaches and workflows emphasize in how simulation is important in the context of model developing. Simulation can help researchers understand how the model behaves in a controlled setting and can be used to stress the model in different ways before it is exposed to any real data. This improved understanding could be beneficial in epidemiological models, specially when dealing with COVID-19. Unfortunately, few researchers perform any simulations. We present a simulation algorithm that implements a simple agent-based model for disease transmission that works with a standard compartment epidemiological model for COVID-19. Our algorithm can be applied in different parameterizations to reflect several plausible epidemic scenarios. Additionally, we also model how social media information in the form of daily symptom mentions can be incorporate into COVID-19 epidemiological models. We test our social media COVID-19 model with two experiments. The first using simulated data from our agent-based simulation algorithm and the second with real data using a machine learning tweet classifier to identify tweets that mention symptoms from noise. Our results shows how a COVID-19 model can be (1) used to incorporate social media data and (2) assessed and evaluated with simulated and real data.
\end{abstract}

% keywords can be removed
\keywords{COVID \and Bayesian Modeling \and Epidemiology \and Simulation \and Social Media \and Twitter}

\section{Introduction}
% I like a 5-paragraph introduction, you can overrule if you want to.

Modern approaches to Bayesian modeling emphasize the importance of developing a model before exposing it to actual data but few researchers actually bother doing it, e.g., \cite{STRINGHINI2020313, ZHANG2020793, RODA2020271,10.1001/jama.2020.17022,kontis2020magnitude,NIEHUS2020803} did not report any sort of simulation or data generating process (DGP) step in their analysis. Simulation of DGPs are especially important for observational studies where fitting a model to data is trivial given the raw curve fitting power of modern techniques so developing against simulations attempts to somewhat separate the model being evaluated from its eventual application to actual data. While lacking the power of a randomized control trial (RCT), a model that performs well across a range of plausible simulations increases the confidence that the fit to actual data is robust and usable for important \citep{IOANNIDIS2020} tasks like estimating future trends for both observed and unobserved variables. 

A secondary benefit of studying simulations is to estimate the impact of model features against simulations that exercise those variables. It is quite easy to determine if a varying rate of infection over time is: 1) recoverable from a simulation that does so; 2) how accurately can the model recover the actual parameter values; and 3) all parameters are available from the simulating DGP whether observed or not. These features quickly identify how the model performs in ways unavailable with real data. Another concern addressed with simulations is the fact that even simple compartment models degrade into chaotic systems under reasonable seeming assumptions such as time varying infection rates \citep{barrientosChaoticDynamicsSeasonally2017}. 

The flow of presentation is as follows:
\begin{enumerate}
    \item We provide background on COVID-19 modeling with an emphases on Bayesian approaches (Section \ref{sec:epimodel});
    \item We present simulation algorithm, available for download, that implements a simple agent-based model for disease transmission that works with a standard compartment model. The simulation is exercised through a single setting of plausible parameterizations (Section \ref{sec:experiments});
    \item We fit a Bayesian compartment epidemiological model to simulated data and compare internal model states, e.g. the compartment populations, to those of the DPG simulation (Section \ref{sec:simulated});
    \item We also fit the model to actual data from Brazil and discuss the challenges to our use case (Section \ref{sec:running_brazil}); and
    \item We summarize our conclusions and also address limitations and opportunities for future studies (Section \ref{sec:conclusion}).
\end{enumerate}

% \section{Social Media}
% \label{sec:socialmedia}

% \subsection{Twitter}
% \label{subsec:twitter}

\section{Simulating an Epidemiological Model}
\label{sec:epimodel}

Compartmental models, also called population-based models, are used to model the dynamics of a infectious disease in a population scale. Those models simplify the complex reality of an epidemic by subdividing the total population into homogeneous groups, called compartments. Individuals within the same compartment are considered to be in the same state regarding the progression of the disease. Compartmental models originated in the beginning of the 20th century with the Susceptible-Infectious-Recovered (SIR) model \citep{kermack1927contribution} which splits the population in three time-dependent compartments: the susceptible, the infected (and infectious), and the recovered (and not infectious) compartments. When a susceptible individual comes into contact with an infectious individual, the former can become infected for some time, and then recover and become immune.

Some infectious diseases are fatal, so in order to differentiates between recovered and deceased, the Susceptible-Infectious-Recovered-Deceased (SIRD) model \citep{bailey1975mathematical} was developed. Since COVID-19 can quickly overcome a nation's health system by overloading the need for intensive care unit (ICU) beds \citep{pintonetoMathematicalModelCOVID192021}, we found the need to include a state that represents terminally-ill patients. Our core model includes also a $T$ state for terminally-ill individuals who have been infected and will unfortunately become deceased. The acronym then becomes Susceptible-Infectious-Recovered-Terminally-ill-Deceased (SIRTD) although we will do experiments with simpler and more complex models.

The dynamics of SIRTD are governed by a system of ordinary differential equations (ODE):

\begin{align}
    \frac{dS}{dt} &= -\beta  S \frac{I}{N} \label{eq:ode_dS}\\
    \frac{dI}{dt} &= \beta  S  \frac{I}{N} - \frac{1}{d_I}  I \label{eq:ode_dI}\\
    \frac{dR}{dt} &= \frac{1}{d_I} I \left( 1 - \omega \right) \label{eq:ode_dR}\\
    \frac{dT}{dt} &= \frac{1}{d_I} I \omega - \frac{1}{d_T} T \label{eq:ode_dT}\\
    \frac{dD}{dt} &= \frac{1}{d_T} T \label{eq:ode_dD}
\end{align}

where:

\begin{itemize}
    \item $S(t)$ is the number of people susceptible to becoming infected (no immunity);
    \item $I(t)$ is the number of people currently infected (and infectious);
    \item $T(t)$ is the number of terminally ill individuals who have been infected and will die;
    \item $R(t)$ is the number of removed people (either dead or we assume they remain immune indefinitely);
    \item $D(t)$ is the number of recovered people that unfortunately died;
    \item $N = S(t) + I(t) + R(t) + T(t) + D(t)$ is the constant total number of individuals in the population;
    \item $\beta$ is the constant rate of contacts between individuals per unit time that are sufficient to lead to transmission if one of the individuals is infectious and the other is susceptible;
    \item $\omega$ is constant death rate of recovered individuals;
    \item $d_I$ is the mean time for which individuals are infectious; and
    \item $d_T$  is the mean time for which individuals are terminally-ill.
\end{itemize}

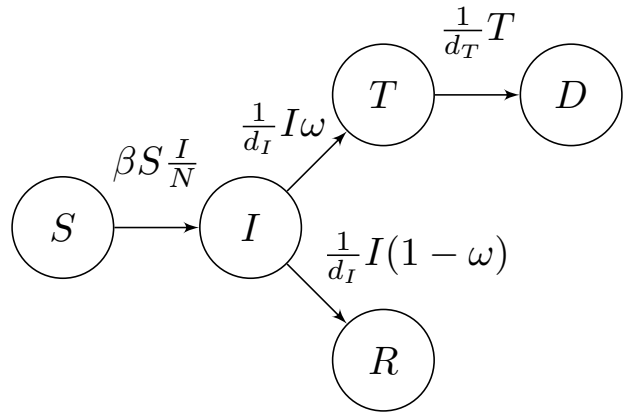
\begin{figure}[!ht]
    \centering
    \scalebox{1.5}{    \begin{tikzpicture}[node distance=0.75cm,auto,>=latex',every node/.append style={align=center},state/.style={draw, circle, minimum size=0.9cm}]
        \node [state] (S) {$S$};
        \node [state, right=of S] (I) {$I$};
        \node [state, above right=of I] (T) {$T$};
        \node [state, below right=of I] (R) {$R$};
        \node [state, right=of T] (D) {$D$};

        \path[->, auto=false] (S) edge node {$\beta S \frac{I}{N}$ \\[2.5em]} (I);
        \path[->, auto=false] (I) edge node {$\hspace{-1.75em} \frac{1}{d_I} I \omega$ \\[1.0em]} (T);
        \path[->, auto=false] (I) edge node {$\hspace{5em} \frac{1}{d_I} I (1 - \omega)$ \\[1.2em]} (R);
        \path[->, auto=false] (T) edge node {$\frac{1}{d_T} T$ \\[2.5em]} (D);
    \end{tikzpicture}}
    \caption{SIRTD Compartmental Model}
    \label{fig:model}
\end{figure}

Susceptible individuals (state $S$) will randomly get in contact with infected individuals (state $I$) and, consequently from this contact, become infected with rate $\beta$ (equation \ref{eq:ode_dS}). Once the susceptible individual becomes infected, he/she can infect other susceptible individuals by random encounters and stays infected/infectious for an average of $d_I$ days (equation \ref{eq:ode_dI}). Infected individuals can recover (state $R$) with probability $1-\omega$ (equation \ref{eq:ode_dR}) or become terminally-ill (state $T$) with probability $\omega$ (equation \ref{eq:ode_dT}). Finally, terminally-ill individuals will eventually decease (state $D$) in an average of $d_T$ days (equation \ref{eq:ode_dD}). The model can also be represented in an directed acyclic graph (DAG) in figure \ref{fig:model}.

The SIRTD model has several assumptions. First, it assumes that population $N$ is constant. Second, every state is populated by homogeneous individuals, i.e., no differences in demographics, social characteristics or health-related variables. Third, the model assumes a random mixing of the population, susceptible are in contact with infectious only governed by chance alone. Fourth, infected will become infectious (they can spread the disease) and will either recover or become terminally-ill. Fifth, infected will also, during the time that they remain infected, potentially infecting susceptible, i.e., no self-quarantine or isolation measures are taken. Finally, recovered are forever immune.

\subsection{The Role of Simulation in Epidemiological Models}
\label{sec:epimodel_simulation}

One of our main contributions is to propose and execute a simulation-driven modeling. We agree with \cite{IOANNIDIS2020} that it is important to "careful modeling of predictive distributions ... and continuously reappraising models based on their validated performance" is essential in epidemiological modeling. There is lack of attention to simulation in the recent epidemiology literature, specially related to COVID-19. Searching in \textit{Scopus} for epidemiology models for COVID-19 in the top peer-reviewed journals, we find that most do not analyze their models with regard to how well they perform in a controlled DGP simulation. For instance, \cite{STRINGHINI2020313, ZHANG2020793, RODA2020271,10.1001/jama.2020.17022,kontis2020magnitude,NIEHUS2020803} all used a Bayesian model and did only inference using the likelihood conditional on data. Despite that, we found some evidence of simulation and care for how the proposed model perform in a controlled setting \citep{Braunereabd9338, biology9050097}

\subsection{Simulating an Epidemiological Model with Social Media Data}
\label{sec:epimodel_socialmedia}
% Twitter Stuff

COVID-19 modeling appears amenable to heterogeneous information sources informing modeling and many ideas have been explored. This work was inspired by the CoDatMo's Liverpool Model \citep{LiverpoolCodatmo}\footnote{The data and code for CoDatMo's Liverpool model can be found here: \url{https://codatmo.github.io/Liverpool/}.} that combined 111 calls reporting symptoms to health authorities with weekly death data in a sophisticated SEEIIRTTD model. Liverpool also kindly provided Twitter data in Portuguese filtered for symptoms so we credit them with setting the form of our model and information sources. However the richness and quality of data in emerging countries can be quite different which raises issues around how complex a modeling solution is possible for Brazil. One goal of this paper, currently unachieved, is an assessment of the benefits of model complexity as we compare performance of SIR, SIRD, SIRTD and other models with DGPs that are themselves of varying complexity. 

Despite several attempts of real-time pandemic monitoring and forecast we found no literature that incorporate social media data into epidemiological models. It is quite common to use epidemiological models for real-time monitoring and forecasting of COVID-19 dynamics but without any social media data \citep{birrellRealtimeNowcastingForecasting2020, jersakova2021bayesian, altmejd2020nowcasting, Schneble_2020, hawryluk2021gaussian, diloro2020nowcasting, wang2020spatiotemporal, stoner2020powerful}.

Studies that did use social media data in our explorations were preoccupied with network analyses \citep{mattei2021italian, esquirol2020characterizing, chiresaire2020characterizing, cruickshank2020characterizing}, semantic meaning \citep{chopra2021mining, Wicke_2021, kruspe2020crosslanguage}, depression and suicide \citep{cortes2020covid19}, fake news \citep{yang2021covid19, shahi2020exploratory, singh2020look}, companies' challenges \citep{patuelli2021firms}, drug mentions \citep{tekumalla2020characterizing}, and privacy issues \citep{dev2020discussing}.

Furthermore, some studies tried to extract information regarding COVID-19 dynamics from social media but without incorporating this information into epidemiological models. \cite{zong2020extracting} presented an annotated corpus of
7,500 tweets for COVID-19 events demonstrating the possibility of accurately identifying COVID-19 events in Twitter but with no extensions to COVID-19 dynamics or modeling efforts. In the same line, \cite{Kaushal_2020} trained a natural language processing (NLP) deep learning model to detect COVID-19 related events from Twitter, such as individuals who recently contracted the virus, someone with symptoms who were denied testing and believed remedies against the infection. A similar approach was done by \cite{santosh2020detecting} in detecting symptoms in Twitter. There is also efforts to combine official COVID-19 data from national and international authorities with social media data \citep{pu2020challenges}. One interesting breakthrough came from \cite{Gencoglu_2020} which used causal modeling to discover and quantify causal relationships between pandemic characteristics and Twitter activity as well as public sentiment and showed that twitter data can successfully capture the epidemiological domain knowledge.

We could use both social media and also mobility data in epidemiological model. Mobility data can be easily obtained, for example \cite{avelarWeeklyBayesianModelling2021} Google's mobility data and a Bayesian epidemiological model to predict deaths. Combining social media data with mobility data also presents some issues. One major obstacle is that a small fraction of tweets are geotagged and some of them have inaccurate location data \citep{huang2020twitter, Porcher_2021}.

To address those gaps, we devised a SIRTD model that uses symptom mentions in social media to better infer and predict the number of infected individuals (state $I$). Our intent is to demonstrate how social media data, specially symptoms mentions, could enhance simple epidemiological models. In the next section we demonstrate our experiments using both simulated and real data from Brazil.

\section{Experiments}
\label{sec:experiments}

We conducted two experiments. The first experiment was with simulated data where configuring parameters were randomly from reasonable ranges and then used to generate data for model fitting. For this preliminary work we ran a single simulation with our SIRTD model.  The second experiment was with real data from Brazil in 2020 where we again run with ou SIRTD model.

We followed the Bayesian workflow for disease transmission modeling by \cite{grinsztajnBayesianWorkflowDisease2021} in which we \textit{build a model}, \textit{fit the model}, \textit{criticize}, and \textit{repeat}. This cycle is also similar to the Bayesian workflow proposed by \cite{gelmanBayesianWorkflow2020} that includes \textit{three} steps of \textit{model building}, \textit{inference}, and \textit{model checking/improvement}, along with the \textit{comparison of different models}.

For all experiments we used \texttt{Stan} \citep{carpenterStanProbabilisticProgramming2017}: a Bayesian probabilistic programming language for specifying complex statistical models and performing inference using Markov Chain Monte Carlo (MCMC). All the data, source code and \texttt{Stan} models can be found on a \texttt{GitHub} repository\footnote{\url{https://github.com/codatmo/dataGeneratingProcess1}.}. The ODE system of equations described in equations \ref{eq:ode_dS}, \ref{eq:ode_dI}, \ref{eq:ode_dR}, \ref{eq:ode_dT}, \ref{eq:ode_dD} were implemented and solved by a 4th/5th order Runge-Kutta method \citep{irseles2008numericalanalysis} using the Dormand-Prince algorithm \citep{dormandFamilyEmbeddedRungeKutta1980}\footnote{\texttt{Stan} implements ODE solvers from \texttt{Boost} (library \texttt{Odeint}) \citep{ahnertOdeintSolvingOrdinary2011} and exposes as a set of two functions \texttt{ode\_rk45} and \texttt{ode\_rk45\_tol}, for automatic or additional control parameters for the solver, respectively.} with relative tolerance and absolute tolerance of \expnumber{1}{-6} and maximum number of steps $h = \text{\expnumber{1}{4}}$.

The model can be specified as following. First, the prior distributions specifications. The constant rate of infection $\beta$ is sampled from a normal distribution constrained to positive values (equation \ref{eq:prior_beta}) with mean $\mu_\beta$ and standard deviation $\sigma_\beta$. The constant death rate of recovered individuals $\omega$ is sampled from a beta distribution (equation \ref{eq:prior_omega}) with parameters $\alpha_\omega$ representing the number of people that unfortunately will become terminally-ill and deceased and $\beta_\omega$ representing the number of people that will recover from the disease. The mean time for which individuals are either infectious or terminally-ill, $d_I$ and $d_T$, are both sampled from a normal distribution constrained to positive values (equations \ref{eq:prior_dI} and \ref{eq:prior_dT}) with means $\mu_{d_I}, \mu_{d_T}$ and standard deviation $\sigma_{d_I}, \sigma_{d_T}$ respectively. The proportion of infected people who will tweet daily about his/her symptoms, $\text{Proportion Tweets}$, while being in state $I$ is sampled from a flat prior distribution for proportions as a beta distribution (equation \ref{eq:prior_prop_twitter}).

The model has the following likelihood specifications. Both daily counts of tweets regarding symptoms and cumulative deaths counts are distributed as negative binomial distribution\footnote{we used the alternative negative binomial parameterization which has mean (i.e., location) parameter and a parameter that controls over-dispersion relative to the square of the mean (i.e., scale parameter): \texttt{Stan}'s \texttt{neg\_binomial\_2}.}. For cumulative death counts (equation \ref{eq:likelihood_deaths}), the location parameter is the number of individuals in state $D$ (solved by \texttt{Stan}'s ODE solver) and the precision parameter $\phi$ which follows an exponential distribution with rate parameter $\lambda_\phi$ (equation \ref{eq:prior_phi}). For daily counts of tweets regarding symptoms (equation \ref{eq:likelihood_tweets}), the location parameter is number of individuals in state $I$ (also solved by \texttt{Stan}'s ODE solver) multiplied by the proportion of infected people who will tweet daily about his/her symptoms, $\text{Proportion Tweets}$, while being in state $I$; and the precision parameter $\phi_{\text{tweets}}$ which follows an exponential distribution with rate parameter $\lambda_{\phi_{\text{tweets}}}$ (equation \ref{eq:prior_phi_twitter}).

\begin{align}
    \beta &\sim \text{Normal}^+(\mu_\beta, \sigma_\beta)                            \label{eq:prior_beta}\\
    \omega &\sim \text{Beta}(\alpha_\omega, \beta_\omega)                           \label{eq:prior_omega}\\
    d_I &\sim \text{Normal}^+(\mu_{d_I}, \sigma_{d_I})                              \label{eq:prior_dI}\\
    d_T &\sim \text{Normal}^+(\mu_{d_T}, \sigma_{d_T})                              \label{eq:prior_dT}\\
    \text{Proportion Tweets} &\sim \text{Beta}(1,1)                                 \label{eq:prior_prop_twitter}\\
    \phi &\sim \text{Exponential}(\lambda_\phi)                                     \label{eq:prior_phi}\\
    \phi_{\text{tweets}} &\sim \text{Exponential}(\lambda_{\phi_{\text{tweets}}})   \label{eq:prior_phi_twitter}\\
    \text{Deceased} &\sim \text{Negative Binomial}\left(\text{state $D$}, \frac{1}{\phi}\right) \label{eq:likelihood_deaths}\\
    \text{Tweets} &\sim \text{Negative Binomial}\left(\text{state $I$} \cdot \text{Proportion Tweets}, \frac{1}{\phi_{\text{tweets}}}\right) \label{eq:likelihood_tweets}
\end{align}

In all of our experiments, we set the priors for the model as similar priors that are used in some COVID-19 epidemiological models \citep{LiverpoolCodatmo}:
\begin{itemize}
    \setlength\itemsep{-0.25em}
    \item[] $\beta \sim \text{Normal}^+ (2, 1)$;
    \item[] $\omega \sim \text{Normal}^+ (0.4, 0.5)$;
    \item[] $\lambda \sim \text{Beta}+ (1, 2)$;
    \item[] $d_I \sim \text{Normal}^+ (7, 2)$;
    \item[] $d_T \sim \text{Normal}^+ (10, 2)$;
    \item[] $\phi \sim \text{Exponential} (5)$; and
    \item[] $\phi_{\text{tweets}} \sim \text{Exponential} (5)$.
\end{itemize}

For all of our sampling, we mostly used \texttt{Stan}'s defaults settings. This translates to MCMC sampling using Hamiltonian Monte Carlo (HMC) \citep{neal2011mcmc} and No-U-Turn-Sampling (NUTS) \citep{hoffman2014no} with 4 separated chains, each having 2,000 iterations and the first 1,000 (half of the total iterations) being discarded as \textit{warm-up}
and the last 1,000 being used as samples from the underlying Markov chain. We took care to set specific random number generator seeds to make our results reproducible. We also used default's parameters for the NUTS HMC sampler, which means the target Metropolis acceptance rate is 80\% (\texttt{adapt\_delta = 0.8}) and the cap on the depth of the trees that it evaluates during each iteration is $2^{10}$ (\texttt{max\_treedepth = 10}).

Our computing environment uses \texttt{R} version 4.1.0 \citep{Rlang}, \texttt{Stan} version 2.27.0 \citep{carpenterStanProbabilisticProgramming2017}, \texttt{CmdStanR} version 0.4.0 \citep{cmdstanr}.

\subsection{Simulated Data}
\label{sec:simulated}

Our simulation closely mirrors the structure of the SIRTD model described above in part to help debug and better understand the dynamics of the models being fit. As a result the simulated data most likely is too easy for the model to recover but we anticipate complicating the simulation in later versions of this work to break the near isomorphism between the model and the simulated data generating process (DGP). 

Algorithm \ref{algo_SIRTD_tweets} is the pseudo-code representation of our agent-based simulation. Starting from everyday we reset the twitter count and then start to simulate each individual independently depending on what compartment the individual is in the current day of the epidemic simulation. If an individual is in the infected $I$ compartment, the individual will tweet about his or hers symptoms with probability $\lambda$ and will have $C$ daily contacts with other individuals from a population $N$. If one of those contacts is an individual in the susceptible $S$ compartment, then the susceptible individual will become infected with probability $\beta$. Everyday an infected individual will have a change to recover and leave the infected $I$ compartment with probability $\frac{1}{d_I}$. The infected individual, then can leave either to the terminally-ill $T$ compartment with probability $\omega$ or to the recovered $R$ compartment with probability $1 - \omega$. Finally, if the individual is in the terminally-ill $T$ compartment, the individual will leave to the deceased $D$ compartment with probability $\frac{1}{d_T}$

    \begin{algorithm}[H]
    \SetAlgoLined
    \SetKwArray{SimData}{sim\_data}
    \SetKwArray{Beta}{$\boldsymbol{\beta}$}
    \SetKwData{CurrentState}{current\_state}
    \SetKwData{Tweets}{tweets}
    \KwIn{
    \begin{itemize}
        \setlength\itemsep{-0.25em}
        \item[] $N$: population size
        \item[] $t$: number of days
        \item[] $C$: mean daily contacts between infected and susceptible
        \item[] $\beta$: infection rate
        %\item[] \beta{$t$}$: daily varying infection rate % for weekly varying stuff in the future
        \item[] $\omega$: fatality rate
        \item[] $\lambda$: daily probability of infected individuals in $I$ compartment tweeting about their symptoms
        \item[] $d_I$: mean of dwell time in $I$ compartment
        \item[] $d_T$: mean of dwell time in $T$ compartment
        \item[] $I_0$: number of $I$ in initial time
    \end{itemize}}
    \KwResult{\SimData{$t$, $6$}: Simulated data for SIRTD model}
    initialization\;
    \SimData{$1$, $6$} $\leftarrow$ [$S = N - I_0$, $I = I_0$, $R$ = 0, $T$ = 0, $D$ = 0, \Tweets = 0]\;
    \For{$i \leftarrow 2$ \KwTo $t$ }{
        \Tweets $\leftarrow 0 $\;
        \For{$p \leftarrow 1$ \KwTo $N$}{
            \If{$p \in I$}{
                \Tweets += Bernoulli$(\lambda)$\;
                \If{Bernoulli$\left(\frac{1}{d_I}\right)$}{
                    $I$ -= 1\;
                    \lIf{Bernoulli$(\omega)$}{
                        $T$ += 1}
                    \lElse{
                        $R$ += 1}}
                    \For{$p_{other} \in$ Sample($C \in N$)}{
                        \If{$p_{other} \in S$ and Bernoulli$(\beta)$}{
                            $I$ +=1\;
                            $S$ -= 1\;}}
                }
            \If{$p \in T$}{
                \If{Bernoulli$\left(\frac{1}{d_T}\right)$}{
                    $T$ -= 1\;
                    $D$ += 1\;}}
        }
        \SimData{$t$, $6$} $\leftarrow$ [$S$, $I$, $R$, $T$, $D$, \Tweets]\;
    }
    \caption{SIRTD with Tweets Simulation}\label{algo_SIRTD_tweets}
    \end{algorithm}

The chief benefit to simulations is that it forces one to confront the details of the model from a generation perspective independent of the model being created to characterize, in this case COVID-19, the phenomenon of study. It is our opinion that just running a Bayesian model generatively does not satisfy the intent nor yield the benefit of a fully specified DGP as done with a prior predictive check. Exercising the likelihood in this way yields little additional knowledge other than the fact that the priors can be recovered with success. 

Our agent-based model is nearly isomorphic in parameterization and execution to the SIRTD model we fit it with but even this level of simulation provided insights. In an agent based framework there has to be more thought given about how one agent infects another. For example a person who is $I$ must come in contact with people who are $S$ and those interactions have to be $\beta$ infectious on average from one day to the next. There are also no fractional people in our simulated world so our solution was to posit some number of interactions $C$ per day for $I$ people with $S$ people with $\frac{\beta}{C}$ chance of being infected. $C$ obviously will not stay constant presumably over their time as $I$ but we ignore that, as we do the possibility that $\beta$ is greater than one. 

For our simulated data, actual parameter values were set as following:
\begin{itemize}
    \setlength\itemsep{-0.25em}
    \item[] $N = 10,000$;
    \item[] $t = 70$;
    \item[] $C = 10$;
    \item[] $\beta = 0.3$;
    \item[] $\omega = 0.1$;
    \item[] $\lambda = 0.2$;
    \item[] $d_I = 7$;
    \item[] $d_T = 10$; and
    \item[] $I_0 = 10$.
\end{itemize}

In Table \ref{tbl:sim_results}, we show the parameters recovered by our SIRTD model. The model could recover all true values for the simulated parameters. The sampling had no divergences and also with good convergence estimates, i.e. all \texttt{rhat}s are below or equal to 1.01.

\begin{table}[!ht]
    \centering
    
    \begin{tabular}{|c | c c c c c c c c c|} 
    \hline
    variable & mean & median & sd & mad & q5 & q95 & rhat & ess\_bulk & ess\_tail\\
    \hline
    $\beta$ & 0.24 & 0.24 & 0.00 & 0.00 & 0.23 & 0.24 & 1.00 & 1560.24 & 2030.97\\
    $\omega$ & 0.10 & 0.10 & 0.00 & 0.00 & 0.09 & 0.11 & 1.00 & 1843.85 & 2149.88\\
    $\lambda$ & 0.12 & 0.12 & 0.01 & 0.01 & 0.11 & 0.14 & 1.00 & 1507.30 & 1940.91\\
    $d_I$ & 10.57 & 10.53 & 0.69 & 0.67 & 9.52 & 11.78 & 1.00 & 1496.19 & 1836.27\\
    $d_T$ & 11.00 & 10.96 & 1.25 & 1.25 & 8.95 & 13.11 & 1.00 & 1972.81 & 2390.61\\
    \hline
    \end{tabular}

    \caption{Model Summary for Simulated Data}\label{tbl:sim_results}
\end{table}

\begin{figure}[!htp]
    \centering
    \includegraphics[width=0.75\textwidth]{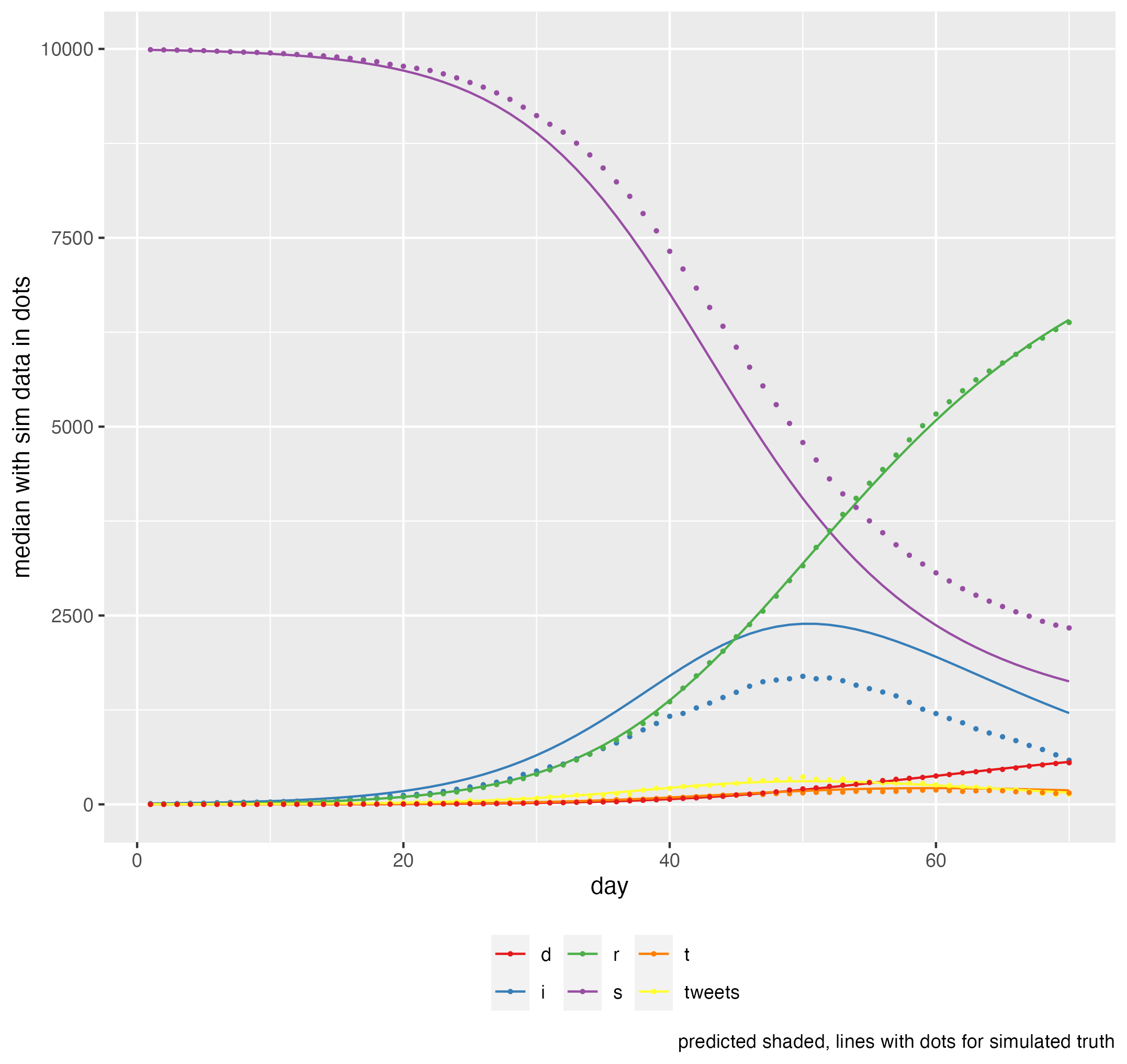}
    \caption{Simulated Data and Model Estimates}
    \label{fig:fit_simulated}
\end{figure}

Also in Figure \ref{fig:fit_simulated} we display the simulated truth for all the compartments in the SIRTD simulated data in dots and the mean generated prediction values as lines, which shows visually how the model can closely replicate trends after inferring the parameter values.

\subsection{Brazil's COVID-19 Data}
\label{sec:brazil}

\begin{figure}[!htp]
    \centering
    \includegraphics[width=0.75\textwidth]{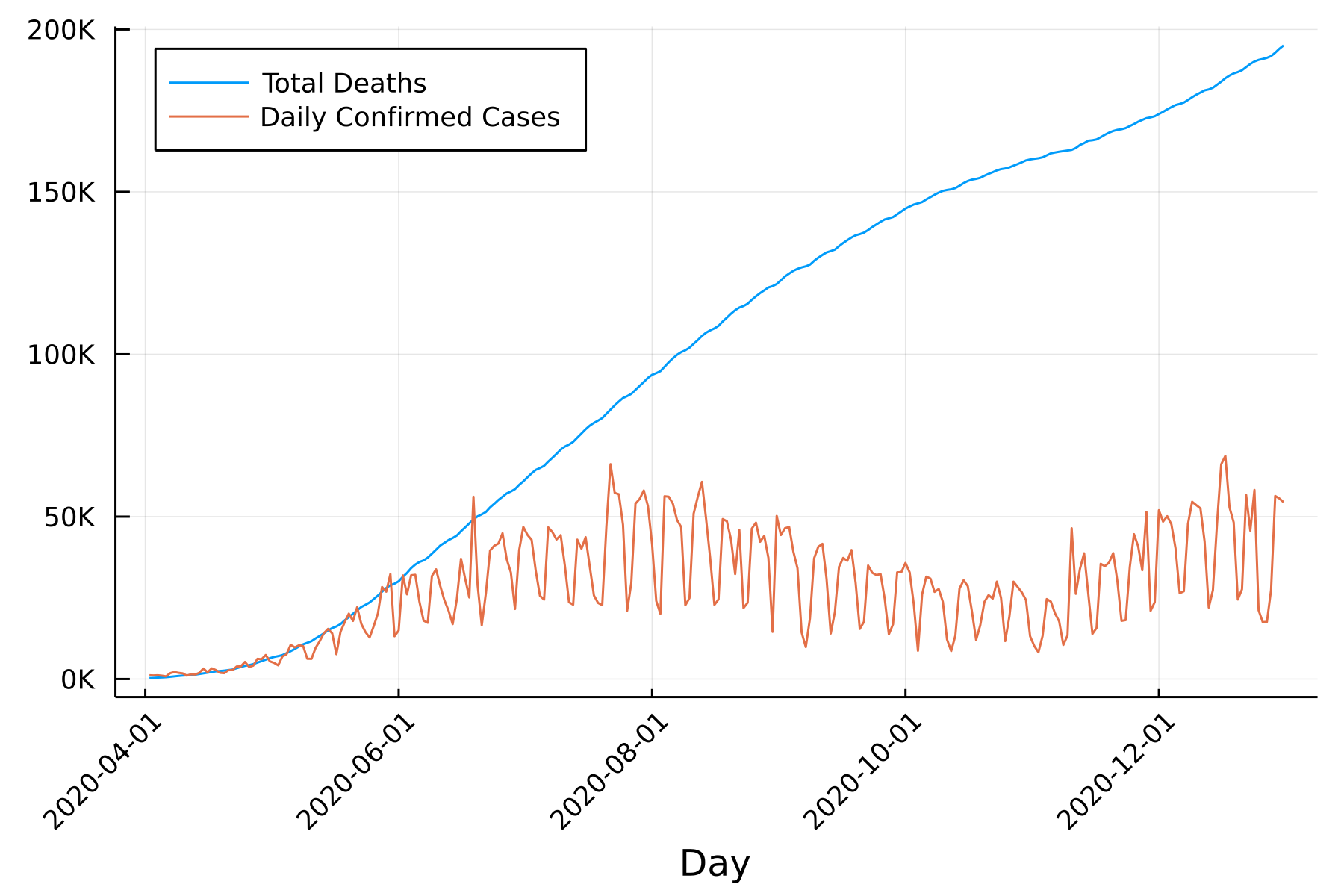}
    \caption{COVID-19 Daily Infected and Total Deaths in 2020}
    \label{fig:brazil_2020}
\end{figure}

In Brazil, events such as Flu Syndrome (FS) and Severe Acute Respiratory Illness (SARI) are countrywide notified since the beginning of the SARS-CoV-2 pandemic. Flu Syndrome cases that seek the health system for COVID-19 testing are registered in the \textit{e-SUS Notifica} information system, and those who are hospitalized or die due to SARI are notified in the Epidemiological Surveillance Information System of \textit{Sivep-Gripe}. The SARI surveillance, implemented in 2009 with the advent of the influenza A (H1N1) pdm09 virus pandemic, is carried out in all public or private hospitals in the country that have capacity to provide assistance to cases of SARI \citep{GuiaCovid}. We used data from the \textit{e-SUS Notifica} but only restricted our analyses to the year 2020. In Figure \ref{fig:brazil_2020} we show the daily confirmed cases by PCR-positive results in red and total cumulative daily deaths in blue.

% Flu Syndrome occurrences are under-reported, as they depend on the health system's testing capacity and the population's opportunity to seek for diagnosis; thus, the information generated by this system has an unknown representativeness, which limits its use for modeling purposes and even for other frequentist analyses.

% The SARI case information system, in turn, is more reliable, as it accounts for all hospitalized cases and deaths with additional information about these cases, such as identification and sociodemographic data, symptom onset, comorbidities and risk factors, hospitalization data such as location, ventilatory support and the use of Intensive Care Unit (ICU), confirmation/discard of the etiological agent, case evolution (discharge or death), evolution date, among others.

% Thus, there is great reliability in the use of data from this system for different analyses, not to mention that these records are available, in real time, in open databases, such as openDataSUS, from the Ministry of Health of Brazil, with the exclusion of only some variables in order not to allow the identification of cases (patient's name, mother's name, address, etc.), due to confidentiality concerns.

\begin{figure}[!htp]
    \centering
    \includegraphics[width=0.75\textwidth]{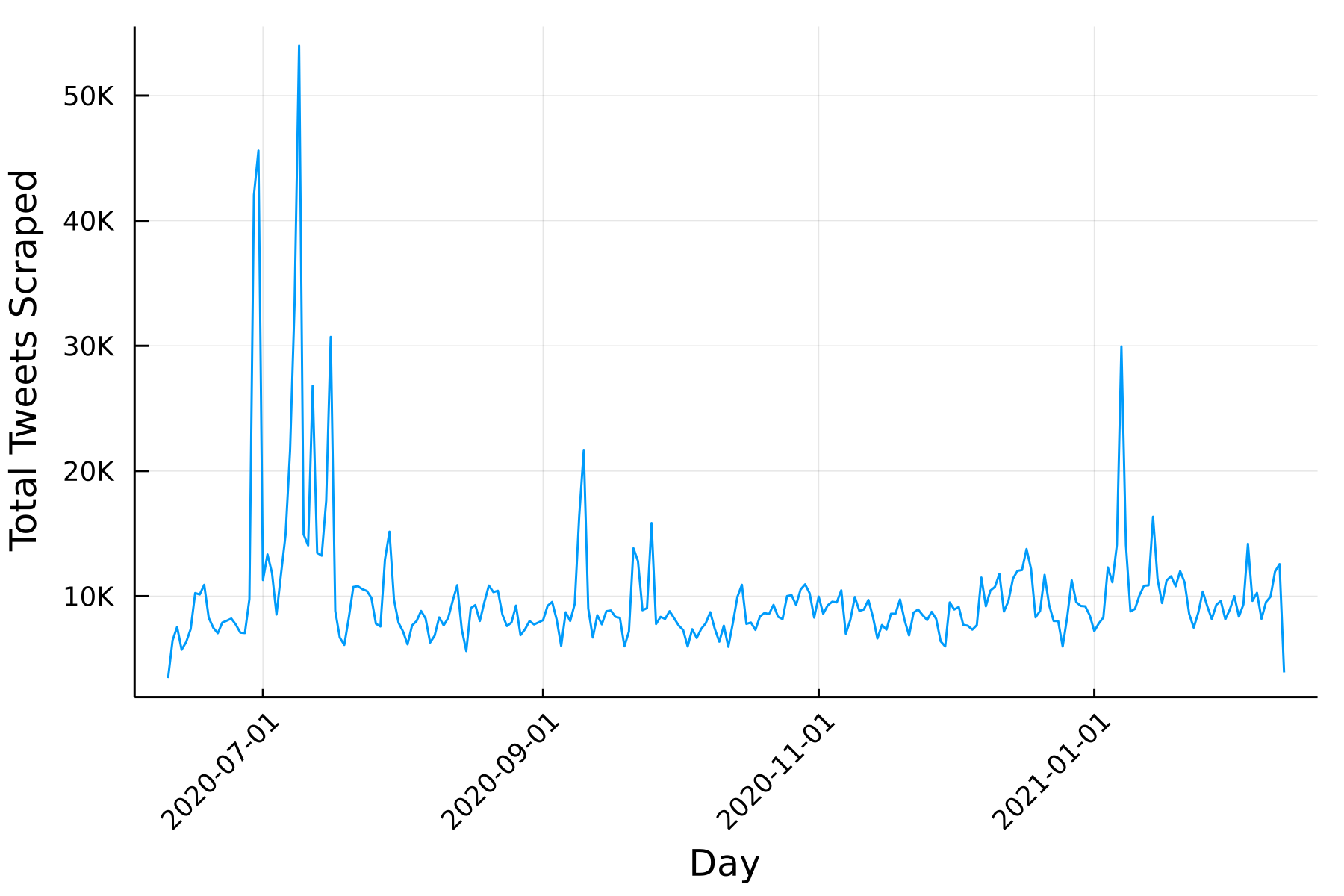}
    \caption{Tweets Scraped}
    \label{fig:tweets_scraped}
\end{figure}

To address those gaps, we devised a list of 56 keywords including signs and symptoms compatible with COVID-19, such as flu-like symptoms, body pain, fever, cough, runny nose, anosmia, respiratory distress and other related terms; that were used to webscrape
% version 2 with 12mi tweets from the beginning of the COVID-19 pandemic in Brazil on February, 25th 2020 to December, 31st 2020.
2,042,775 tweets from June, 10th 2020 to December, 31st 2020 (see Figure \ref{fig:tweets_scraped}).

\begin{figure}[!htp]
    \centering
    \includegraphics[width=0.75\textwidth]{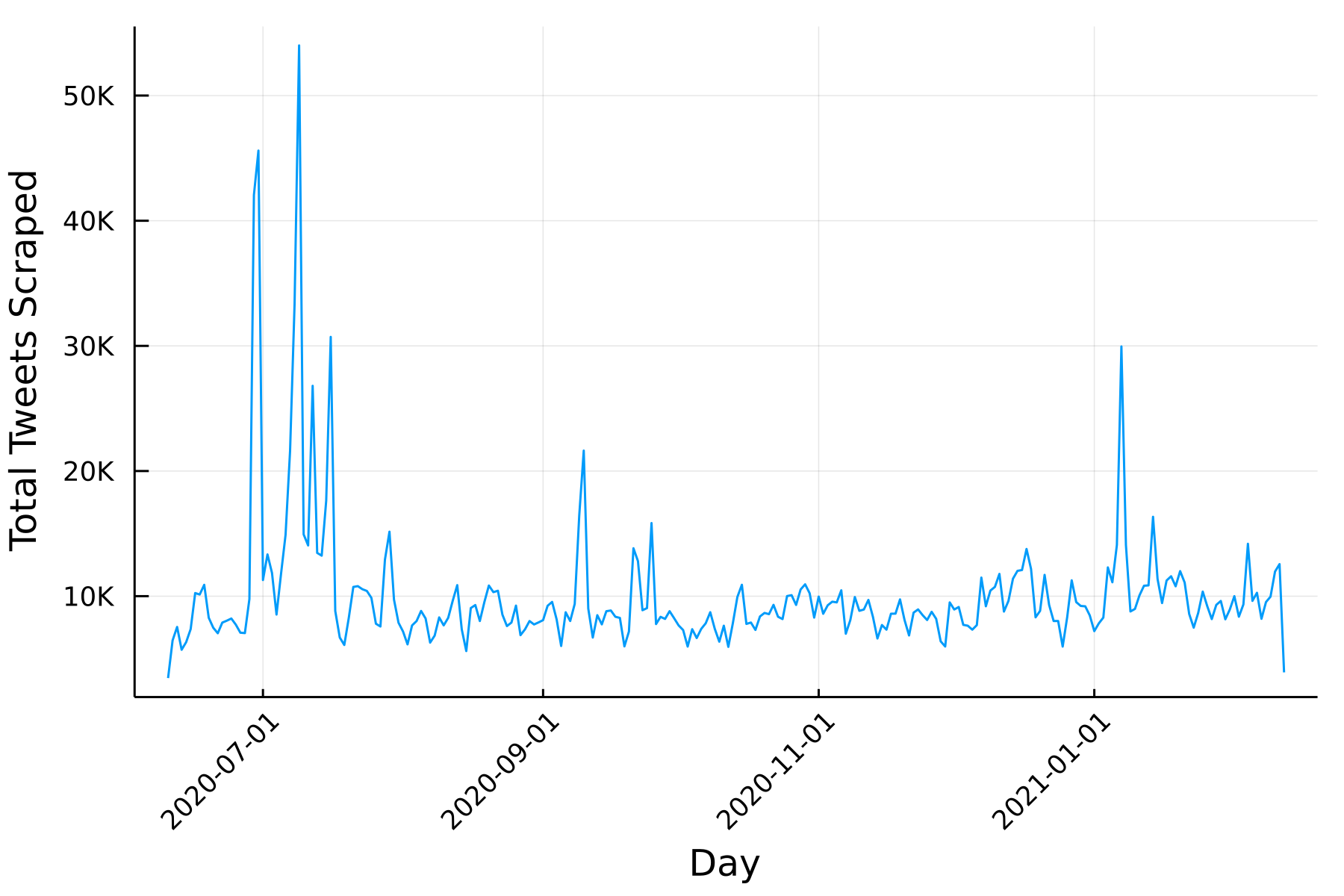}
    \caption{Daily Tweets Predicted with Signal}
    \label{fig:tweets_predicted}
\end{figure}

We annotated 9,600 tweets with binary labels indicating 0 for noise and 1 for signal regarding the mention of symptoms either by the user or by someone that the user knows. Those labeled tweets were used to train a term frequency–inverse document frequency (TF-IDF) \citep{salton1988term} Random Forest classifier in scikit-learn \citep{scikit-learn}\footnote{the data for the Brazilian Portuguese tweet classifier can be found on GitHub: \url{https://github.com/codatmo/Brazil-Tweet-Classifier}.}. We achieved achieved 90\% accuracy in the test set (80/20 split) (see Table \ref{tbl:clasifier_metrics}). The trained classified was then used to predict the remaining unlabeled tweets either with noise (0) or signal (1). We used the aggregated daily counts to generate our twitter symptom mention time series used in the Brazil's COVID-19 model inferences and predictions (see Figure \ref{fig:tweets_predicted}).

\begin{table}[!ht]
    \centering
    
    \begin{tabular}{|c | c c c|} 
    \hline
    label & precision & recall & f1-score \\
    \hline
    0 & 0.94 & 0.93 & 0.93 \\
    1 & 0.75 & 0.80 & 0.78 \\
    \hline
    \end{tabular}

    \caption{Classification Metrics for the Twitter Symptoms Classifier}\label{tbl:clasifier_metrics}
\end{table}

Some studies used data from Brazil's social media in epidemiological models of dengue. \cite{albinati2017enhancement} used Twitter data to improve epidemiological models for predicting dengue incidence in real time in Brazil. \cite{Souza_2019} detected spatial clusters of dengue risk using Twitter data in two Brazilian cities with more than 1 million inhabitants and the highest dengue incidence rates in 2015. \cite{Souza_2015} developed a latent shared-component generative model to predict dengue outbreaks in Brazilian urban areas, also using data collected from Twitter.

\subsection{Running Brazil Data}
\label{sec:running_brazil}

We also ran our SIRTD model for the Brazilian real data. Since we have only tweets from June 10th 2020 onwards we used official deaths and confirmed data from this date onwards. In a future version of this preprint we will use twitter data since the beginning of the COVID-19 pandemic in Brazil (February 25th of 2020). For the Brazilian data, we used population values from the official last available government data (year 2019). For the initial individuals counts in the SIRTD compartments, we set $I_0$ to be the number of PCR-positive COVID cases for 10th June 2020, $R_0$ as cumulative total of PCR-positive COVID cases in 10th June 2020, $D_0$ as the cumulative deaths in 10th June 2020 and $T_0 = 0$. We subtracted from population the initials $I_0$, $R_0$, $T_0$ and $D_0$ to get the initial susceptible number $S_0$.

Since, we cannot compare the real infection rate $\beta$ and real death rate because of under-reporting in Brazilian data, we cannot compare our model estimated parameters with the ground truth. In Table \ref{tbl:brazil_results}, we show the parameters recovered by our SIRTD model. The model could recover all parameters without any divergence with good convergence estimates, i.e. all \texttt{rhat}s are below or equal to 1.01.

\begin{table}[!ht]
    \centering
    
    \begin{tabular}{|c | c c c c c c c c c|} 
    \hline
    variable & mean & median & sd & mad & q5 & q95 & rhat & ess\_bulk & ess\_tail\\
    \hline
    $\beta$ & 0.14 & 0.13 & 0.03 & 0.03 & 0.01 & 0.29 & 1.01 & 297.49 & 448.45\\
    $\omega$ & 0.29 & 0.28 & 0.06 & 0.06 & 0.20 & 0.39 & 1.01 & 292.22 & 452.46\\
    $\lambda$ & 0.11 & 0.11 & 0.00 & 0.00 & 0.10 & 0.12 & 1.00 & 570.88 & 709.18\\
    $d_I$ & 8.47 & 8.34 & 1.61 & 1.62 & 5.99 & 11.31 & 1.01 & 298.39 & 470.20\\
    $d_T$ & 1.99 & 1.99 & 0.46 & 0.45 & 1.26 & 2.75 & 1.00 & 591.52 & 556.67\\
     \hline
     \end{tabular}

    \caption{Model Summary for Brazil Data}\label{tbl:brazil_results}
\end{table}

\begin{figure}[!htp]
    \centering
    \includegraphics[width=0.75\textwidth]{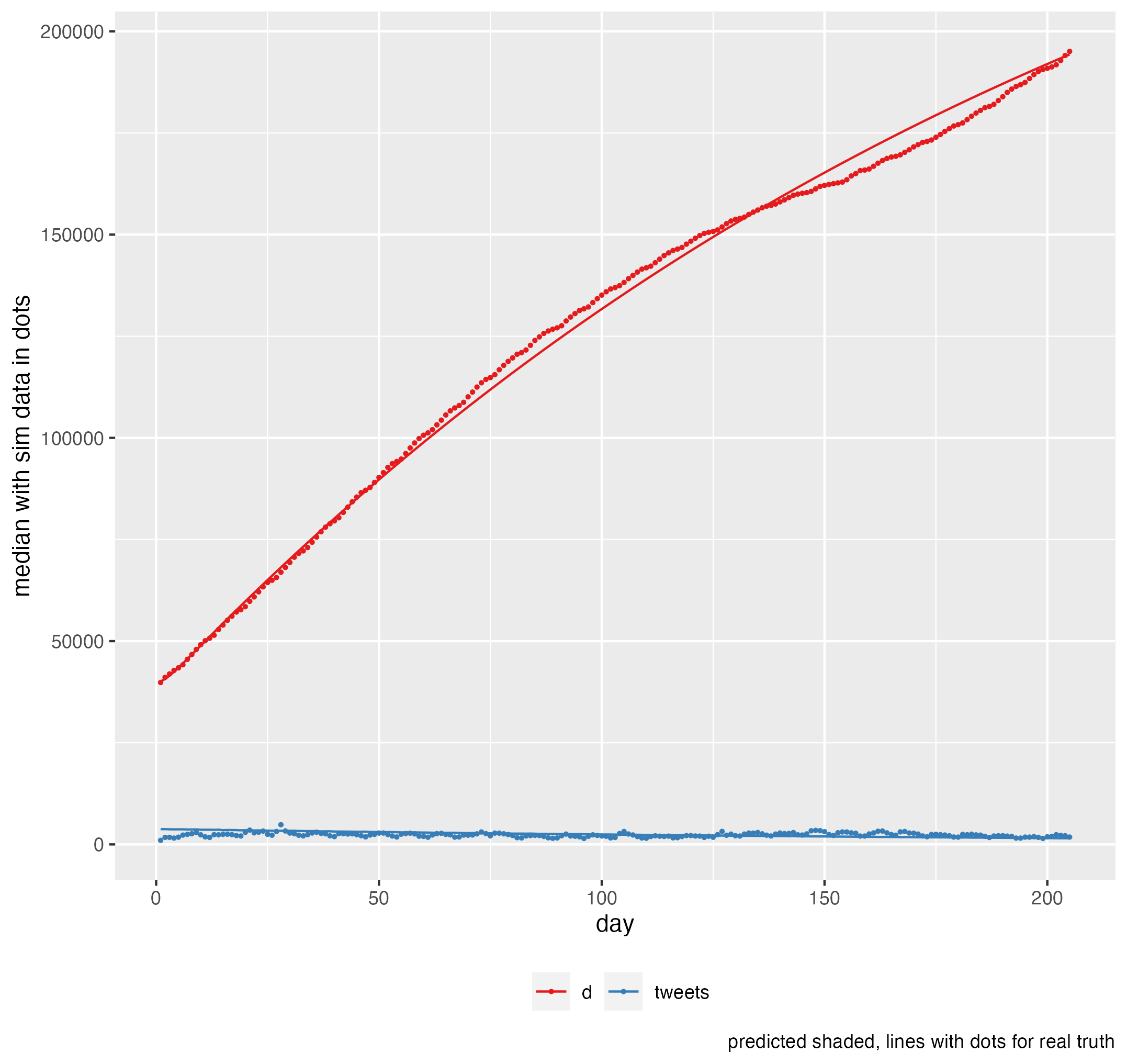}
    \caption{Brazil Data and Model Estimates}
    \label{fig:fit_brazil}
\end{figure}

Also in Figure \ref{fig:fit_brazil} we show the ground truth as dots and the mean generated prediction values as lines, which shows visually how the model could generate accurate predictions after inferring the parameter values.

\section{Conclusion}
\label{sec:conclusion}

One of our contributions is to demonstrate how epidemiological models, specifically in the case of COVID-19 pandemic, can be better comprehended and in turn become more robust. For example, in June 2021, Brazil reached the mark of 500,000 deaths and 18 million accumulated cases of COVID-19 \citep{PainelCovid}. This has impacted the availability of inpatient beds and other assistance resources in a large part of the country. Thus, the development of reliable predictive epidemiological models could help federated entities better plan of assistance to critical cases.

Some limitations should be mentioned: (1) the data was not analyzed considering sociodemographic differences (gender, age group, place of residence, opportunity to access health care, etc.), we did not account for heterogeneous transmission and mortality by age group like \cite{hauserEstimationSARSCoV2Mortality2020}; (2) in most real-world situations, the infection rate $\beta$ varies over time, we did not account for this and we modeled $\beta$ as a constante parameter over time; (e) different control measures were adopted by distinct municipal, state and federal authorities, at different times, in an attempt to contain the disease, besides there is also political biases, we have not incorporate heterogeneous measures by different government authorities in our model; (4) it was not possible to disaggregate the model to state and municipal levels, due to the lack of reliable geotagged social media data; and (5) real data from infected (and infectious) individuals were not used to verify our models that used real data from Brazil, due to under-reporting of mild cases, given the low capacity for population testing in the Brazil. We encorage future studies to address those limitations.

\bibliographystyle{unsrtnat}
\bibliography{references}  

\begin{thebibliography}{62}
\providecommand{\natexlab}[1]{#1}
\providecommand{\url}[1]{\texttt{#1}}
\expandafter\ifx\csname urlstyle\endcsname\relax
  \providecommand{\doi}[1]{doi: #1}\else
  \providecommand{\doi}{doi: \begingroup \urlstyle{rm}\Url}\fi

\bibitem[Stringhini et~al.(2020)Stringhini, Wisniak, Piumatti, Azman, Lauer,
  Baysson, {De Ridder}, Petrovic, Schrempft, Marcus, Yerly, {Arm Vernez},
  Keiser, Hurst, Posfay-Barbe, Trono, Pittet, Gétaz, Chappuis, Eckerle,
  Vuilleumier, Meyer, Flahault, Kaiser, and Guessous]{STRINGHINI2020313}
Silvia Stringhini, Ania Wisniak, Giovanni Piumatti, Andrew~S Azman, Stephen~A
  Lauer, Hélène Baysson, David {De Ridder}, Dusan Petrovic, Stephanie
  Schrempft, Kailing Marcus, Sabine Yerly, Isabelle {Arm Vernez}, Olivia
  Keiser, Samia Hurst, Klara~M Posfay-Barbe, Didier Trono, Didier Pittet,
  Laurent Gétaz, François Chappuis, Isabella Eckerle, Nicolas Vuilleumier,
  Benjamin Meyer, Antoine Flahault, Laurent Kaiser, and Idris Guessous.
\newblock Seroprevalence of anti-sars-cov-2 igg antibodies in geneva,
  switzerland (serocov-pop): a population-based study.
\newblock \emph{The Lancet}, 396\penalty0 (10247):\penalty0 313--319, 2020.
\newblock ISSN 0140-6736.
\newblock \doi{https://doi.org/10.1016/S0140-6736(20)31304-0}.
\newblock URL
  \url{https://www.sciencedirect.com/science/article/pii/S0140673620313040}.

\bibitem[Zhang et~al.(2020)Zhang, Litvinova, Wang, Wang, Deng, Chen, Li, Zheng,
  Yi, Chen, Wu, Liang, Wang, Yang, Sun, Longini, Halloran, Wu, Cowling, Merler,
  Viboud, Vespignani, Ajelli, and Yu]{ZHANG2020793}
Juanjuan Zhang, Maria Litvinova, Wei Wang, Yan Wang, Xiaowei Deng, Xinghui
  Chen, Mei Li, Wen Zheng, Lan Yi, Xinhua Chen, Qianhui Wu, Yuxia Liang, Xiling
  Wang, Juan Yang, Kaiyuan Sun, Ira~M Longini, M~Elizabeth Halloran, Peng Wu,
  Benjamin~J Cowling, Stefano Merler, Cecile Viboud, Alessandro Vespignani,
  Marco Ajelli, and Hongjie Yu.
\newblock Evolving epidemiology and transmission dynamics of coronavirus
  disease 2019 outside hubei province, china: a descriptive and modelling
  study.
\newblock \emph{The Lancet Infectious Diseases}, 20\penalty0 (7):\penalty0
  793--802, 2020.
\newblock ISSN 1473-3099.
\newblock \doi{https://doi.org/10.1016/S1473-3099(20)30230-9}.
\newblock URL
  \url{https://www.sciencedirect.com/science/article/pii/S1473309920302309}.

\bibitem[Roda et~al.(2020)Roda, Varughese, Han, and Li]{RODA2020271}
Weston~C. Roda, Marie~B. Varughese, Donglin Han, and Michael~Y. Li.
\newblock Why is it difficult to accurately predict the covid-19 epidemic?
\newblock \emph{Infectious Disease Modelling}, 5:\penalty0 271--281, 2020.
\newblock ISSN 2468-0427.
\newblock \doi{https://doi.org/10.1016/j.idm.2020.03.001}.
\newblock URL
  \url{https://www.sciencedirect.com/science/article/pii/S2468042720300075}.

\bibitem[for~the REMAP-CAP~Investigators(2020)]{10.1001/jama.2020.17022}
The Writing~Committee for~the REMAP-CAP~Investigators.
\newblock {Effect of Hydrocortisone on Mortality and Organ Support in Patients
  With Severe COVID-19: The REMAP-CAP COVID-19 Corticosteroid Domain Randomized
  Clinical Trial}.
\newblock \emph{JAMA}, 324\penalty0 (13):\penalty0 1317--1329, 10 2020.
\newblock ISSN 0098-7484.
\newblock \doi{10.1001/jama.2020.17022}.
\newblock URL \url{https://doi.org/10.1001/jama.2020.17022}.

\bibitem[Kontis et~al.(2020)Kontis, Bennett, Rashid, Parks, Pearson-Stuttard,
  Guillot, Asaria, Zhou, Battaglini, Corsetti, et~al.]{kontis2020magnitude}
Vasilis Kontis, James~E Bennett, Theo Rashid, Robbie~M Parks, Jonathan
  Pearson-Stuttard, Michel Guillot, Perviz Asaria, Bin Zhou, Marco Battaglini,
  Gianni Corsetti, et~al.
\newblock Magnitude, demographics and dynamics of the effect of the first wave
  of the covid-19 pandemic on all-cause mortality in 21 industrialized
  countries.
\newblock \emph{Nature medicine}, pages 1--10, 2020.

\bibitem[Niehus et~al.(2020)Niehus, {De Salazar}, Taylor, and
  Lipsitch]{NIEHUS2020803}
Rene Niehus, Pablo~M {De Salazar}, Aimee~R Taylor, and Marc Lipsitch.
\newblock Using observational data to quantify bias of traveller-derived
  covid-19 prevalence estimates in wuhan, china.
\newblock \emph{The Lancet Infectious Diseases}, 20\penalty0 (7):\penalty0
  803--808, 2020.
\newblock ISSN 1473-3099.
\newblock \doi{https://doi.org/10.1016/S1473-3099(20)30229-2}.
\newblock URL
  \url{https://www.sciencedirect.com/science/article/pii/S1473309920302292}.

\bibitem[Ioannidis et~al.(2020)Ioannidis, Cripps, and Tanner]{IOANNIDIS2020}
John~P.A. Ioannidis, Sally Cripps, and Martin~A. Tanner.
\newblock Forecasting for covid-19 has failed.
\newblock \emph{International Journal of Forecasting}, 2020.
\newblock ISSN 0169-2070.
\newblock \doi{https://doi.org/10.1016/j.ijforecast.2020.08.004}.
\newblock URL
  \url{https://www.sciencedirect.com/science/article/pii/S0169207020301199}.

\bibitem[Barrientos et~al.(2017)Barrientos, Rodr{\'i}guez, and
  {Ruiz-Herrera}]{barrientosChaoticDynamicsSeasonally2017}
Pablo~G. Barrientos, J.~{\'A}ngel Rodr{\'i}guez, and Alfonso {Ruiz-Herrera}.
\newblock Chaotic dynamics in the seasonally forced {{SIR}} epidemic model.
\newblock \emph{Journal of Mathematical Biology}, 75\penalty0 (6):\penalty0
  1655--1668, December 2017.
\newblock ISSN 1432-1416.
\newblock \doi{10.1007/s00285-017-1130-9}.

\bibitem[Kermack and McKendrick(1927)]{kermack1927contribution}
William~Ogilvy Kermack and Anderson~G McKendrick.
\newblock A contribution to the mathematical theory of epidemics.
\newblock \emph{Proceedings of the royal society of london. Series A,
  Containing papers of a mathematical and physical character}, 115\penalty0
  (772):\penalty0 700--721, 1927.

\bibitem[Bailey et~al.(1975)]{bailey1975mathematical}
Norman~TJ Bailey et~al.
\newblock \emph{The mathematical theory of infectious diseases and its
  applications}.
\newblock Charles Griffin \& Company Ltd, 5a Crendon Street, High Wycombe,
  Bucks HP13 6LE., 1975.

\bibitem[Pinto~Neto et~al.(2021)Pinto~Neto, Kennedy, Reis, Wang, Brizzi,
  Zambrano, {de Souza}, Pedroso, {de Mello Pedreiro}, {de Matos Brizzi},
  Abinader, and Z{\^a}ngaro]{pintonetoMathematicalModelCOVID192021}
Osmar Pinto~Neto, Deanna~M. Kennedy, Jos{\'e}~Clark Reis, Yiyu Wang, Ana
  Carolina~Brisola Brizzi, Gustavo~Jos{\'e} Zambrano, Joabe~Marcos {de Souza},
  Wellington Pedroso, Rodrigo~Cunha {de Mello Pedreiro}, Bruno {de Matos
  Brizzi}, Ellysson~Oliveira Abinader, and Renato~Amaro Z{\^a}ngaro.
\newblock Mathematical model of {{COVID}}-19 intervention scenarios for {{S\~ao
  Paulo}}\textemdash{{Brazil}}.
\newblock \emph{Nature Communications}, 12\penalty0 (1):\penalty0 418, January
  2021.
\newblock ISSN 2041-1723.
\newblock \doi{10.1038/s41467-020-20687-y}.

\bibitem[Brauner et~al.(2021)Brauner, Mindermann, Sharma, Johnston, Salvatier,
  Gaven{\v c}iak, Stephenson, Leech, Altman, Mikulik, Norman, Monrad,
  Besiroglu, Ge, Hartwick, Teh, Chindelevitch, Gal, and
  Kulveit]{Braunereabd9338}
Jan~M. Brauner, S{\"o}ren Mindermann, Mrinank Sharma, David Johnston, John
  Salvatier, Tom{\'a}{\v s} Gaven{\v c}iak, Anna~B. Stephenson, Gavin Leech,
  George Altman, Vladimir Mikulik, Alexander~John Norman, Joshua~Teperowski
  Monrad, Tamay Besiroglu, Hong Ge, Meghan~A. Hartwick, Yee~Whye Teh, Leonid
  Chindelevitch, Yarin Gal, and Jan Kulveit.
\newblock Inferring the effectiveness of government interventions against
  covid-19.
\newblock \emph{Science}, 371\penalty0 (6531), 2021.
\newblock ISSN 0036-8075.
\newblock \doi{10.1126/science.abd9338}.
\newblock URL \url{https://science.sciencemag.org/content/371/6531/eabd9338}.

\bibitem[Roques et~al.(2020)Roques, Klein, Papaïx, Sar, and
  Soubeyrand]{biology9050097}
Lionel Roques, Etienne~K Klein, Julien Papaïx, Antoine Sar, and Samuel
  Soubeyrand.
\newblock Using early data to estimate the actual infection fatality ratio from
  covid-19 in france.
\newblock \emph{Biology}, 9\penalty0 (5), 2020.
\newblock ISSN 2079-7737.
\newblock \doi{10.3390/biology9050097}.
\newblock URL \url{https://www.mdpi.com/2079-7737/9/5/97}.

\bibitem[Moore and Phillips(2021)]{LiverpoolCodatmo}
Robert Moore and Alex Phillips.
\newblock Liverpool covid model.
\newblock \url{https://github.com/codatmo/Liverpool}, 2021.

\bibitem[Birrell et~al.(2020)Birrell, Blake, van Leeuwen, Cell, Gent, and
  Angelis]{birrellRealtimeNowcastingForecasting2020}
Paul Birrell, Joshua Blake, Edwin van Leeuwen, PHE Joint~Modelling Cell, Nick
  Gent, and Daniela~De Angelis.
\newblock Real-time {{Nowcasting}} and {{Forecasting}} of {{COVID}}-19
  {{Dynamics}} in {{England}}: The first wave?
\newblock \emph{medRxiv}, page 2020.08.24.20180737, August 2020.
\newblock \doi{10.1101/2020.08.24.20180737}.

\bibitem[Jersakova et~al.(2021)Jersakova, Lomax, Hetherington, Lehmann,
  Nicholson, Briers, and Holmes]{jersakova2021bayesian}
Radka Jersakova, James Lomax, James Hetherington, Brieuc Lehmann, George
  Nicholson, Marc Briers, and Chris Holmes.
\newblock Bayesian imputation of covid-19 positive test counts for nowcasting
  under reporting lag, 2021.

\bibitem[Altmejd et~al.(2020)Altmejd, Rocklöv, and
  Wallin]{altmejd2020nowcasting}
Adam Altmejd, Joacim Rocklöv, and Jonas Wallin.
\newblock Nowcasting covid-19 statistics reported withdelay: a case-study of
  sweden, 2020.

\bibitem[Schneble et~al.(2020)Schneble, Nicola, Kauermann, and
  Berger]{Schneble_2020}
Marc Schneble, Giacomo~De Nicola, Göran Kauermann, and Ursula Berger.
\newblock Nowcasting fatal {COVID}-19 infections on a regional level in
  germany.
\newblock \emph{Biometrical Journal}, 63\penalty0 (3):\penalty0 471--489, nov
  2020.
\newblock \doi{10.1002/bimj.202000143}.
\newblock URL \url{https://doi.org/10.1002%2Fbimj.202000143}.

\bibitem[Hawryluk et~al.(2021)Hawryluk, Hoeltgebaum, Mishra, Miscouridou,
  Whittaker, Vollmer, Flaxman, Bhatt, and Mellan]{hawryluk2021gaussian}
Iwona Hawryluk, Henrique Hoeltgebaum, Swapnil Mishra, Xenia Miscouridou,
  Ricardo P Schnekenberand~Charles Whittaker, Michaela Vollmer, Seth Flaxman,
  Samir Bhatt, and Thomas~A Mellan.
\newblock Gaussian process nowcasting: Application to covid-19 mortality
  reporting, 2021.

\bibitem[Loro et~al.(2020)Loro, Divino, Farcomeni, Lasinio, Lovison, Maruotti,
  and Mingione]{diloro2020nowcasting}
Pierfrancesco Alaimo~Di Loro, Fabio Divino, Alessio Farcomeni, Giovanna~Jona
  Lasinio, Gianfranco Lovison, Antonello Maruotti, and Marco Mingione.
\newblock Nowcasting covid-19 incidence indicators during the italian first
  outbreak, 2020.

\bibitem[Wang et~al.(2020)Wang, Wang, Gao, Li, Yu, Kim, Wang, and
  ZhilinGu]{wang2020spatiotemporal}
Li~Wang, Guannan Wang, Lei Gao, Xinyi Li, Shan Yu, Myungjin Kim, Yueying Wang,
  and ZhilinGu.
\newblock Spatiotemporal dynamics, nowcasting and forecasting of covid-19 in
  the united states, 2020.

\bibitem[Stoner et~al.(2020)Stoner, Economou, and Halliday]{stoner2020powerful}
Oliver Stoner, Theo Economou, and Alba Halliday.
\newblock A powerful modelling framework for nowcasting and forecasting
  covid-19 and other diseases, 2020.

\bibitem[Mattei et~al.(2021)Mattei, Caldarelli, Squartini, and
  Saracco]{mattei2021italian}
Mattia Mattei, Guido Caldarelli, Tiziano Squartini, and Fabio Saracco.
\newblock Italian twitter semantic network during the covid-19 epidemic, 2021.

\bibitem[Esquirol et~al.(2020)Esquirol, Prignano, Díaz-Guilera, and
  Cozzo]{esquirol2020characterizing}
Bernat Esquirol, Luce Prignano, Albert Díaz-Guilera, and Emanuele Cozzo.
\newblock Characterizing twitter users behaviour during the spanish covid-19
  first wave, 2020.

\bibitem[Chire-Saire(2020)]{chiresaire2020characterizing}
Josimar~E. Chire-Saire.
\newblock Characterizing twitter interaction during covid-19 pandemic using
  complex networks and text mining, 2020.

\bibitem[Cruickshank and Carley(2020)]{cruickshank2020characterizing}
Iain~J. Cruickshank and Kathleen~M. Carley.
\newblock Characterizing communities of hashtag usage on twitter during the
  2020 covid-19 pandemic by multi-view clustering, 2020.

\bibitem[Chopra et~al.(2021)Chopra, Vashishtha, Pal, Ashima, Tyagi, and
  Sethi]{chopra2021mining}
Harshita Chopra, Aniket Vashishtha, Ridam Pal, Ashima, Ananya Tyagi, and
  Tavpritesh Sethi.
\newblock Mining trends of covid-19 vaccine beliefs on twitter with lexical
  embeddings, 2021.

\bibitem[Wicke and Bolognesi(2021)]{Wicke_2021}
Philipp Wicke and Marianna~M. Bolognesi.
\newblock Covid-19 discourse on twitter: How the topics, sentiments,
  subjectivity, and figurative frames changed over time.
\newblock \emph{Frontiers in Communication}, 6, Mar 2021.
\newblock ISSN 2297-900X.
\newblock \doi{10.3389/fcomm.2021.651997}.
\newblock URL \url{http://dx.doi.org/10.3389/fcomm.2021.651997}.

\bibitem[Kruspe et~al.(2020)Kruspe, Häberle, Kuhn, and
  Zhu]{kruspe2020crosslanguage}
Anna Kruspe, Matthias Häberle, Iona Kuhn, and Xiao~Xiang Zhu.
\newblock Cross-language sentiment analysis of european twitter messages
  duringthe covid-19 pandemic, 2020.

\bibitem[Cortes et~al.(2020)Cortes, Muñoz, Betancur, and
  Toro]{cortes2020covid19}
Santiago Cortes, Juan Muñoz, David Betancur, and Mauricio Toro.
\newblock Covid-19 emotion monitoring as a tool to increase preparedness for
  disease outbreaks in developing regions, 2020.

\bibitem[Yang et~al.(2021)Yang, Pierri, Hui, Axelrod, Torres-Lugo, Bryden, and
  Menczer]{yang2021covid19}
Kai-Cheng Yang, Francesco Pierri, Pik-Mai Hui, David Axelrod, Christopher
  Torres-Lugo, John Bryden, and Filippo Menczer.
\newblock The covid-19 infodemic: Twitter versus facebook, 2021.

\bibitem[Shahi et~al.(2020)Shahi, Dirkson, and Majchrzak]{shahi2020exploratory}
Gautam~Kishore Shahi, Anne Dirkson, and Tim~A. Majchrzak.
\newblock An exploratory study of covid-19 misinformation on twitter, 2020.

\bibitem[Singh et~al.(2020)Singh, Bansal, Bode, Budak, Chi, Padden, Vanarsdall,
  Vraga, and Wang]{singh2020look}
Lisa Singh, Shweta Bansal, Leticia Bode, Ceren Budak, Guangqing Chi, Kornraphop
  Kawintiranoand~Colton Padden, Rebecca Vanarsdall, Emily Vraga, and Yanchen
  Wang.
\newblock A first look at covid-19 information and misinformation sharing on
  twitter, 2020.

\bibitem[Patuelli et~al.(2021)Patuelli, Caldarelli, Lattanzi, and
  Saracco]{patuelli2021firms}
Alessia Patuelli, Guido Caldarelli, Nicola Lattanzi, and Fabio Saracco.
\newblock Firms' challenges and social responsibilities during covid-19: a
  twitter analysis, 2021.

\bibitem[Tekumalla and Banda(2020)]{tekumalla2020characterizing}
Ramya Tekumalla and Juan~M. Banda.
\newblock Characterizing drug mentions in covid-19 twitter chatter, 2020.

\bibitem[Dev(2020)]{dev2020discussing}
Jayati Dev.
\newblock Discussing privacy and surveillance on twitter: A case study of
  covid-19.
\newblock \emph{arXiv preprint arXiv:2006.06815}, 2020.

\bibitem[Zong et~al.(2020)Zong, Baheti, Xu, and Ritter]{zong2020extracting}
Shi Zong, Ashutosh Baheti, Wei Xu, and Alan Ritter.
\newblock Extracting covid-19 events from twitter, 2020.

\bibitem[Kaushal and Vaidhya(2020)]{Kaushal_2020}
Ayush Kaushal and Tejas Vaidhya.
\newblock Winners at w-nut 2020 shared task-3: Leveraging event specific and
  chunk span information for extracting covid entities from tweets.
\newblock \emph{Proceedings of the Sixth Workshop on Noisy User-generated Text
  (W-NUT 2020)}, 2020.
\newblock \doi{10.18653/v1/2020.wnut-1.79}.
\newblock URL \url{http://dx.doi.org/10.18653/v1/2020.wnut-1.79}.

\bibitem[Santosh et~al.(2020)Santosh, Schwartz, Eichstaedt, Ungar, and
  Guntuku]{santosh2020detecting}
Roshan Santosh, H.~Andrew Schwartz, Johannes~C. Eichstaedt, Lyle~H. Ungar, and
  Sharath~C. Guntuku.
\newblock Detecting emerging symptoms of covid-19 using context-based twitter
  embeddings, 2020.

\bibitem[Pu et~al.(2020)Pu, Suprem, and Lima]{pu2020challenges}
Calton Pu, Abhijit Suprem, and Rodrigo~Alves Lima.
\newblock Challenges and opportunities in rapid epidemic information
  propagation with live knowledge aggregation from social media, 2020.

\bibitem[Gencoglu and Gruber(2020)]{Gencoglu_2020}
Oguzhan Gencoglu and Mathias Gruber.
\newblock Causal modeling of twitter activity during covid-19.
\newblock \emph{Computation}, 8\penalty0 (4):\penalty0 85, Sep 2020.
\newblock ISSN 2079-3197.
\newblock \doi{10.3390/computation8040085}.
\newblock URL \url{http://dx.doi.org/10.3390/computation8040085}.

\bibitem[Avelar et~al.(2021)Avelar, Lamb, Tsoka, and
  {Cardoso-Silva}]{avelarWeeklyBayesianModelling2021}
Pedro Henrique da~Costa Avelar, Luis~C. Lamb, Sophia Tsoka, and Jonathan
  {Cardoso-Silva}.
\newblock Weekly {{Bayesian}} modelling strategy to predict deaths by
  {{COVID}}-19: A model and case study for the state of {{Santa Catarina}},
  {{Brazil}}.
\newblock \emph{arXiv:2104.01133 [q-bio, stat]}, April 2021.

\bibitem[Huang et~al.(2020)Huang, Li, Jiang, Li, and Porter]{huang2020twitter}
Xiao Huang, Zhenlong Li, Yuqin Jiang, Xiaoming Li, and Dwayne Porter.
\newblock Twitter, human mobility, and covid-19, 2020.

\bibitem[Porcher and Renault(2021)]{Porcher_2021}
Simon Porcher and Thomas Renault.
\newblock Social distancing beliefs and human mobility: Evidence from twitter.
\newblock \emph{PLOS ONE}, 16\penalty0 (3):\penalty0 e0246949, Mar 2021.
\newblock ISSN 1932-6203.
\newblock \doi{10.1371/journal.pone.0246949}.
\newblock URL \url{http://dx.doi.org/10.1371/journal.pone.0246949}.

\bibitem[Grinsztajn et~al.(2021)Grinsztajn, Semenova, Margossian, and
  Riou]{grinsztajnBayesianWorkflowDisease2021}
L{\'e}o Grinsztajn, Elizaveta Semenova, Charles~C. Margossian, and Julien Riou.
\newblock Bayesian workflow for disease transmission modeling in {{Stan}}.
\newblock \emph{arXiv:2006.02985 [q-bio, stat]}, February 2021.

\bibitem[Gelman et~al.(2020)Gelman, Vehtari, Simpson, Margossian, Carpenter,
  Yao, Kennedy, Gabry, B{\"u}rkner, and Modr{\'a}k]{gelmanBayesianWorkflow2020}
Andrew Gelman, Aki Vehtari, Daniel Simpson, Charles~C. Margossian, Bob
  Carpenter, Yuling Yao, Lauren Kennedy, Jonah Gabry, Paul-Christian
  B{\"u}rkner, and Martin Modr{\'a}k.
\newblock Bayesian {{Workflow}}.
\newblock \emph{arXiv:2011.01808 [stat]}, November 2020.

\bibitem[Carpenter et~al.(2017)Carpenter, Gelman, Hoffman, Lee, Goodrich,
  Betancourt, Brubaker, Guo, Li, and
  Riddell]{carpenterStanProbabilisticProgramming2017}
Bob Carpenter, Andrew Gelman, Matthew~D. Hoffman, Daniel Lee, Ben Goodrich,
  Michael Betancourt, Marcus Brubaker, Jiqiang Guo, Peter Li, and Allen
  Riddell.
\newblock Stan : {{A Probabilistic Programming Language}}.
\newblock \emph{Journal of Statistical Software}, 76\penalty0 (1), 2017.
\newblock ISSN 1548-7660.
\newblock \doi{10.18637/jss.v076.i01}.

\bibitem[Iserles(2008)]{irseles2008numericalanalysis}
Arieh Iserles.
\newblock \emph{A First Course in the Numerical Analysis of Differential
  Equations}.
\newblock {Cambridge University Press}, {USA}, second edition, 2008.
\newblock ISBN 0-521-73490-8.

\bibitem[Dormand and Prince(1980)]{dormandFamilyEmbeddedRungeKutta1980}
J.~R. Dormand and P.~J. Prince.
\newblock A family of embedded {{Runge}}-{{Kutta}} formulae.
\newblock \emph{Journal of Computational and Applied Mathematics}, 6\penalty0
  (1):\penalty0 19--26, March 1980.
\newblock ISSN 0377-0427.
\newblock \doi{10.1016/0771-050X(80)90013-3}.

\bibitem[Ahnert and Mulansky(2011)]{ahnertOdeintSolvingOrdinary2011}
Karsten Ahnert and Mario Mulansky.
\newblock Odeint \textendash{} {{Solving Ordinary Differential Equations}} in
  {{C}}++.
\newblock \emph{AIP Conference Proceedings}, 1389\penalty0 (1):\penalty0
  1586--1589, September 2011.
\newblock ISSN 0094-243X.
\newblock \doi{10.1063/1.3637934}.

\bibitem[Neal(2011)]{neal2011mcmc}
Radford~M Neal.
\newblock {{MCMC}} using {{Hamiltonian}} dynamics.
\newblock In Steve Brooks, Andrew Gelman, Galin~L. Jones, and Xiao-Li Meng,
  editors, \emph{Handbook of Markov Chain Monte Carlo}. Chapman and Hall/CRC,
  2011.

\bibitem[Hoffman and Gelman(2011)]{hoffman2014no}
Matthew~D Hoffman and Andrew Gelman.
\newblock The {{No}}-{{U}}-{{Turn Sampler}}: {{Adaptively Setting Path
  Lengths}} in {{Hamiltonian Monte Carlo}}.
\newblock \emph{Journal of Machine Learning Research}, 15\penalty0
  (1):\penalty0 1593--1623, November 2011.

\bibitem[{R Core Team}(2021)]{Rlang}
{R Core Team}.
\newblock \emph{R: A Language and Environment for Statistical Computing}.
\newblock R Foundation for Statistical Computing, Vienna, Austria, 2021.
\newblock URL \url{https://www.R-project.org/}.

\bibitem[Gabry and Češnovar(2021)]{cmdstanr}
Jonah Gabry and Rok Češnovar.
\newblock \emph{cmdstanr: R Interface to 'CmdStan'}, 2021.
\newblock https://mc-stan.org/cmdstanr, https://discourse.mc-stan.org.

\bibitem[{Ministério da Saúde do Brasil (a)}(2021)]{GuiaCovid}
{Ministério da Saúde do Brasil (a)}.
\newblock Guia de vigilância epidemiológica: Emergência de saúde pública
  de importância nacional pela doença pelo coronavírus 2019 - covid-19
  [recurso eletrônico], 2021.
\newblock URL
  \url{https://www.gov.br/saude/pt-br/coronavirus/publicacoes-tecnicas/guias-e-planos/guia-de-vigilancia-epidemiologica-covid-19/view}.

\bibitem[Salton and Buckley(1988)]{salton1988term}
Gerard Salton and Christopher Buckley.
\newblock Term-weighting approaches in automatic text retrieval.
\newblock \emph{Information processing \& management}, 24\penalty0
  (5):\penalty0 513--523, 1988.

\bibitem[Pedregosa et~al.(2011)Pedregosa, Varoquaux, Gramfort, Michel, Thirion,
  Grisel, Blondel, Prettenhofer, Weiss, Dubourg, Vanderplas, Passos,
  Cournapeau, Brucher, Perrot, and Duchesnay]{scikit-learn}
F~Pedregosa, G~Varoquaux, A~Gramfort, V~Michel, B~Thirion, O~Grisel, M~Blondel,
  P~Prettenhofer, R~Weiss, V~Dubourg, J~Vanderplas, A~Passos, D~Cournapeau,
  M~Brucher, M~Perrot, and E~Duchesnay.
\newblock Scikit-learn: Machine learning in python.
\newblock \emph{Journal of Machine Learning Research}, 12:\penalty0 2825--2830,
  2011.

\bibitem[Albinati et~al.(2017)Albinati, au2, Pappa, Teixeira, and
  Marques-Toledo]{albinati2017enhancement}
Julio Albinati, Wagner Meira~Jr. au2, Gisele~L. Pappa, Mauro Teixeira, and
  Cecilia Marques-Toledo.
\newblock Enhancement of epidemiological models for dengue fever based on
  twitter data, 2017.

\bibitem[Souza et~al.(2019)Souza, Assunção, Oliveira, Neill, and
  Meira]{Souza_2019}
Roberto~C.S.N.P. Souza, Renato~M. Assunção, Derick~M. Oliveira, Daniel~B.
  Neill, and Wagner Meira.
\newblock Where did i get dengue? detecting spatial clusters of infection risk
  with social network data.
\newblock \emph{Spatial and Spatio-temporal Epidemiology}, 29:\penalty0
  163--175, 2019.
\newblock ISSN 1877-5845.
\newblock \doi{https://doi.org/10.1016/j.sste.2018.11.005}.
\newblock URL
  \url{https://www.sciencedirect.com/science/article/pii/S1877584517301715}.

\bibitem[Souza et~al.(2015)Souza, de~Brito, Assunção, and au2]{Souza_2015}
Roberto C. S. N.~P. Souza, Denise E.~F de~Brito, Renato~M. Assunção, and
  Wagner Meira~Jr au2.
\newblock A latent shared-component generative model for real-time disease
  surveillance using twitter data, 2015.

\bibitem[{Ministério da Saúde do Brasil (b)}(2021)]{PainelCovid}
{Ministério da Saúde do Brasil (b)}.
\newblock Painel coronavirus, 2021.
\newblock URL \url{https://covid.saude.gov.br/}.

\bibitem[Hauser et~al.(2020)Hauser, Counotte, Margossian, Konstantinoudis, Low,
  Althaus, and Riou]{hauserEstimationSARSCoV2Mortality2020}
Anthony Hauser, Michel~J. Counotte, Charles~C. Margossian, Garyfallos
  Konstantinoudis, Nicola Low, Christian~L. Althaus, and Julien Riou.
\newblock Estimation of {{SARS}}-{{CoV}}-2 mortality during the early stages of
  an epidemic: {{A}} modeling study in {{Hubei}}, {{China}}, and six regions in
  {{Europe}}.
\newblock \emph{PLOS Medicine}, 17\penalty0 (7):\penalty0 e1003189, July 2020.
\newblock ISSN 1549-1676.
\newblock \doi{10.1371/journal.pmed.1003189}.

\end{thebibliography}

\end{document}